\documentclass[aps,amsfonts,amsmath,prd,nofootinbib,preprint,tightenlines]{revtex4-1}
\newcommand{\beq}{\begin{equation}}
\newcommand{\eeq}{\end{equation}}
\newcommand{\bea}{\begin{eqnarray}}
\newcommand{\eea}{\end{eqnarray}}

\usepackage{epsfig,bbm}

\begin{document}

\title{Large parallel cosmic string simulations:\\
 New results on loop production}

\author{Jose J.  Blanco-Pillado}
\author{Ken D. Olum}
\author{Benjamin Shlaer}
\affiliation{Institute of Cosmology, Department of Physics and Astronomy,\\ 
Tufts University, Medford, MA 02155, USA}

\begin{abstract}
Using a new parallel computing technique, we have run the largest
cosmic string simulations ever performed. Our results confirm the
existence of a long transient period where a non-scaling distribution
of small loops is produced at lengths depending on the initial
correlation scale. As time passes, this initial population gives way
to the true scaling regime, where loops of size approximately equal to
one-twentieth the horizon distance become a significant component. We
observe similar behavior in matter and radiation eras, as well as in
flat space.  In the matter era, the scaling population of large loops
becomes the dominant component; we expect this to eventually happen in
the other eras as well.
\end{abstract}


\maketitle
\thispagestyle{empty}
\section{Introduction}
\setcounter{page}{1}
The formation and evolution of field theoretic cosmic strings has been
extensively studied for the past thirty years (See \cite{Alex-book} and
references therein). The idea \cite{Witten} that superstrings
could be stretched to cosmological scales by the expansion of the universe 
has recently \cite{Sarangi-Tye,Dvali-Vilenkin,Copeland-Myers-Polchinski}
revived the interest in cosmic string networks. The
existence of a cosmological network of strings may yield
observational signatures that can be detected with current or planned experiments, 
giving us the opportunity to probe new
physics at tremendous energy scales. This has motivated the study of many
aspects of cosmic strings.

Early work on cosmic strings focused on the simple and generic
observational predictions arising from their kinematic gravitational effects,
namely searching for their imprint on the cosmic microwave background via the
Kaiser-Stebbins effect \cite{Kaiser}, or by looking for the telltale identical pair of
images of an astrophysical object being lensed by a cosmic string
\cite{gravitational-lenses}. These effects are enhanced by
increasing the energy scale of the string, and can therefore be used 
to place an upper bound on the dimensionless string 
tension $G\mu$ \cite{bounds-from-CMB,bounds-from-CMB-field-theory,Bevis-et-al}. 
Another observational signature
predicted by the evolution of a cosmic string network is the existence
of a stochastic background of gravitational waves \cite{gravitational-radiation}
emitted by the oscillating string loops which continually break off the 
network. This is an important effect since it allows the loops to shrink and decay, 
preventing them from becoming a dominant contribution to the
energy budget of our observable universe. Particular features of the
string evolution, such as cusps and kinks, create a focused burst rather than
stochastic gravitational wave emission, and so allow much lighter strings to produce a
detectable intermittent signal \cite{gravitational-radiation-bursts}. Additionally, cosmic strings
can also produce other forms of radiation, either due to the partial
annihilation of the string itself in regions of high curvature
\cite{particle-radiation,particle-radiation-from-field-theory,our-field-theory,Moore-field-theory}, 
or due to the their coupling to some other
nongravitational degree of freedom such as an axion
\cite{axionic-radiation}, a dilaton \cite{dilaton-radiation}, or some
other light field \cite{light-radiation}. Radiation of this type has
been studied in connection with cosmic ray physics, where we can use the
observational bounds on these fluxes to place limits on the parameters
of the cosmic string models \cite{CR-from-strings}.

However, it is clearly necessary to understand the
statistical properties of the cosmic string network in detail before
we can obtain robust predictions of observational
signatures.  Characterizing the essential properties of the
network throughout its evolution has been approached in several
different ways. Early work on this subject studied the properties of
the network with analytical methods
\cite{Vilenkin-81,Kibble-85,Bennett}.  The central thesis
of this work was the approach of the network to a {\it
  scaling solution} whereby the strings contribute a constant
fraction of the energy density of the universe.  The existence of a 
scaling solution is paramount to the viability of cosmic string models, 
since the network would otherwise become the
dominant fraction of energy in the universe, in clear contradiction with
observations.

After the initial impetus from the analytical description of the
scaling solution, people turned to numerical simulation as a technique
to compute the relevant parameters of the string network. Several
groups independently developed codes to evolve a network of Nambu-Goto
equations in an expanding universe
\cite{Albrecht-Turok-1,Albrecht-Turok-2,Bennett-Bouchet-1,Bennett-Bouchet-2,Allen-Shellard}. Perhaps
the most
interesting conclusion of all these papers was that there does exist a
scaling regime for long (``infinite") strings, where the average
distance between the strings, $d(t)$, and the coherence length,
$\xi(t)$, both scale in proportion to the particle horizon distance,
given by $d_h(t) = a \eta$, where $\eta$ is the conformal time and $a$
the scale factor.

\begin{equation}
d(t) \sim \xi(t) \sim d_h(t) \sim t.
\end{equation}

In fact, although the codes were significantly different, they
appeared to be close to a quantitative agreement on the parameters of
the network.  

Further analytical work has been done by several different groups
\cite{Austin-Copeland-Kibble,Martins-Shellard-1,Polchinski-et-al}, but
the necessity of large numerical simulations in understanding the
properties of the string network over a large range of scales has
always remained. The constant improvement of computers and the
introduction of new algorithms allowed for a second generation of
numerical simulations
\cite{Ringeval-1,VOV-05,Martins-Shellard-2,OV-06,VOV-06} of much
larger size. Ref.~\cite{Ringeval-1} studied the loop number
density and loop energy distribution and found an approach to scaling
values of these quantities for loop sizes above a few thousands of the
horizon size.  Then Refs.~\cite{OV-06,VOV-06} studied the loop
production function and found an approach to scaling in that function,
but their results were not compatible with the loop distribution found
by \cite{Ringeval-1}.

Refs.~\cite{bounds-from-CMB-field-theory,Bevis-et-al,particle-radiation-from-field-theory} 
simulated cosmic string loop networks using
lattice field theory, but their results did not agree with the
Nambu-Goto simulations of other groups.  Instead, they found very
significant emission directly from long strings.  Attempts by other
groups to reproduce this effect in field theory simulations
\cite{our-field-theory,Moore-field-theory} did not succeed.

One of the realizations which emerged from all these efforts was that
there are two different time scales associated with the approach of a
network to scaling. The first time scale parametrizes the way long
strings in the network reach a steady state with the correct
macroscopic properties of the scaling solution, and a second, much
longer time scale is associated with the small scale structure of the
network. In other words, the component of the network produced in
loops, as well as the short wavelength spectrum of perturbations on
long strings seem to approach scaling at a slower rate. This makes
their study much more difficult in numerical simulations, where one is
limited by computational resources. This has motivated us to develop a
code based on new parallel simulation techniques
\cite{numerical-techniques-paper} which allows us to run simulations
with dynamic range an order of magnitude larger than those previously
reported in the literature.  The results of these simulations will be
described in a series of papers. Here we focus on the loop production
function.

With a larger dynamic range, it becomes possible to clearly
distinguish behavior at two relevant length scales, namely
the ``smallest scales'' (i.e., the simulation resolution,
the gravitational backreaction scale, or in our case the scale of
initial conditions), and the scale set by the particle horizon. There 
is an ongoing controversy as to which scale is important in 
determining the typical size at which loops are produced.  Roughly 
speaking, the claims are that all loops are produced at scales set 
by the particle horizon \cite{Vilenkin-81,Kibble-85}, 
that all loops are produced at the smallest scales 
\cite{Bennett-Bouchet-1,Bennett-Bouchet-2,Allen-Shellard}, or that some mixture of both are 
produced \cite{Polchinski-et-al, Vitaly}.

Our results confirm those of Olum and Vanchurin \cite{OV-06}, and our
new techniques enable us to study loop production in much more detail
and to much later times.  We see a significant fraction of loop
production in a broad, scaling peak reaching downward from about one
twentieth of the horizon scale, along with an initially dominant but
decreasing population at the smallest scales.  This is true in both
the matter and radiation eras, as well as in flat space.  In the
matter era, the horizon-scale population becomes dominant at late
times.  We expect this to happen as well in the radiation and
flat-space cases, but we are not currently able to run long enough
simulations to see whether such regimes develop.  There appear to be
two different peaks in the scaling spectrum of loops at significant
fractions of the horizon size.


\section{Simulation techniques}

The first step of any cosmic string simulation is to establish the
initial distribution of strings. We have used the procedure of
Vachaspati and Vilenkin \cite{VV} to generate the initial
configuration of the network.  This procedure constructs a string
from the square faces that it must pass through.  For each such
face, we choose a random position on that face equally distributed in
a square filling the central 20\% of the face in each direction, and
we connect each position to the next by a straight string segment.
For each such segment, we then choose a velocity $v$ in a random
direction perpendicular to the string with $v^2$ evenly distributed
between 0 and 0.25.  These random perturbations mitigate the lattice
nature of the initial conditions.

The algorithm for the subsequent evolution of Nambu-Goto strings is
based on that presented in \cite{VOV-05}, where the strings are
described by a collection of straight segments linked together to form
kinky long strings and loops.  These piecewise linear strings can be
simulated exactly (up to computational arithmetic error) in flat space
\cite{VOV-05}. In an expanding background one can write the equations
of motion for the string in conformal coordinates so that the causal
structure of Minkowski space is preserved. The resulting equations cannot 
be analytically solved, but they can be well approximated by expanding the
solution to first order in the product of $H$, the expansion rate of
the universe, and the segment size being simulated \cite{OV-06}. This
approximation improves as the simulation progresses since the Hubble
distance grows relative to the size of the segments. 

The other important ingredient in this algorithm is the intercommutation
of string when two pieces of the string worldsheet intersect.  It is
believed that most gauge theory cosmic strings would intercommute
with a probability $P$ of essentially unity. But this may not be true for
fundamental strings in a 4-dimensional compactification,
where the probability of such a process
is suppressed.  In the present paper we will only explore the case
$P=1$. Other cases will be studied in subsequent publications.

The evolution of the string network continually produces loops of all
sizes. It is useful to distinguish between the sub-horizon ``loops"
and the super-horizon ``long strings" which make up the rest of the
network.  In our periodic space, all simulated strings are technically
loops, and so we classify them by defining loops to be those strings
which will never intersect any other string, including themselves.
Due to the nature of our evolution algorithm it would be
computationally intractable to evolve tiny loops, so we must remove
them at some point.  Loops whose physical size is much smaller than
the interstring distance are likely to evolve without intersecting
with any other string. In flat space we expect small loops to fragment
until eventually all the string energy in the loop ends up in
non-self-intersecting trajectories.  It is then safe to log and remove
those loops from the simulation, since they would not have any effect
on the network properties. In an expanding universe the situation is
complicated by the fact that loops are affected by the cosmological
friction, so a non-self-intersecting loop could, in principle, be
destabilized and chop itself up, triggering a new stage of
fragmentation. We expect this effect to be negligible for any loop of
size significantly smaller than the Hubble distance. We have checked
that this is not an important effect by letting non-self-intersecting
loops oscillate a number of times before being removed. The results do
not show any significant deviation in any of these cases.

One of the main goals of this project is to extend the dynamic range
of prior simulations performed using similar algorithms.  Due to the
fact that the simulations exist in a periodic space, the dynamic range
is limited by the time it takes information to propagate across the
box. A possible solution to this problem was implemented in
\cite{VOV-05,VOV-06} by periodically doubling the size of the box of
the simulation. Here we have chosen an alternative way to extend the
running time, by splitting the box into different regions that are
simulated by many different computers, in other words using a parallel
code. This allows us to faithfully simulate cosmic strings in a much
bigger box, and consequently to run to much later times.  We
developed, to this end, a parallel Nambu-Goto string simulation using
the algorithm of \cite{VOV-05,VOV-06,OV-06} and reported on
preliminary results in \cite{Talloires}. This parallel method proved
to be indeed superior to the single processor codes, but it also had
some inefficiencies associated with the very different work loads that
different processors encounter in a simulation of this kind. At late
times, some regions of the simulation volume will have little or no
string, while others will have a great deal.  In such a case, the
power of parallel computing is greatly diminished, since all the other
processors must wait for the one with the most simulation work to do.
Faced with that difficulty we have developed a new parallelization
technique, whereby the spacetime volume is divided into small
4-volumes whose boundaries are lightlike, i.e. they have only initial
and final surfaces.  This permits the individual 4-volumes to run
independently of one another, imposing only that each individual job
must wait for those who provide data for its initial surfaces. This
new method is particularly useful for our type of simulation, and it
has allowed us to increase the dynamic range by almost an order of
magnitude with respect to all previous cosmic string simulations.  A
detailed account of this new technique and its comparison with
conventional parallelization methods is given in
\cite{numerical-techniques-paper}.

Our technique
naturally gives rise to a periodic simulation volume in the form of a
rhombic dodecahedron with opposite faces identified.  A point and its
images form a face-centered cubic lattice.  We will refer to the
constant comoving distance between a point and its closest images as
the size of the simulation volume, $L$.  The comoving volume being
simulated is thus $V = L^3/\sqrt{2}$.  The face-centered cubic lattice
is close-packed, so the ratio of $V$ to $L^3$ is the smallest
possible.  As shown in \cite{OV-06}, the rate of information
propagation through the simulation volume is about half the speed of
light, so in a box of size $L$, we can run a simulation whose duration
in conformal time is also $L$.

We have run simulations in flat space of size 2000 for a conformal
time duration\footnote{We work throughout in units where the speed of
  light is 1.}  2000 in units of the initial cell size of the
Vachaspati-Vilenkin algorithm.  Such a simulation contains at maximum
about 14 billion linear pieces of string simultaneously, and in the
course of evolution creates about 1 trillion segments and produces
about 10 billion loops of string.  If one counts the largest total
amount of data existing at one moment of simulation time, we believe
this ranks among the largest simulations of any kind ever performed.

In the expanding universe we are not able to simulate as large a
volume.  The reason is that as the universe expands, the comoving
length of linear pieces of string decreases (i.e., they are not purely
stretched with the expansion).  Thus, while the density of string in
expanding cases is not very different from the flat case at the same
time, the number of pieces of string to be simulated is much larger,
and so the simulation effort much greater.

In the case of the radiation era, this has limited us so far to
simulations of size 1500.  In the matter era, the expansion is much
more rapid and the problem more severe; the largest simulations
we have so far run had size 500.

Once one has decided to create strings with a Vachaspati-Vilenkin cell
size of 1, one must choose a conformal time $\eta_i$ for the start of
the simulation.  This time determines the horizon distance, which will
be used to define scaling, and in the expanding universe it controls
the Hubble constant, which determines the rate of redshifting and stretching.  We
choose $\eta_i$ so that the initial conditions correspond as closely
as possible to the conditions that one would expect (based on later
stages in the simulation) in a scaling solution at that time.

In the matter era, we found the best initial conformal time to be 4.5.
We can thus run until conformal time 504.5, without contamination from the
periodicity of the volume.  Our dynamic range, defined as the ratio of
final to initial conformal time, is thus 112.  In the radiation era,
we chose initial time 6.  We can thus run until conformal time 1506,
for dynamic range 251.  In flat space, we chose initial time 4.0 and
ran until time 2004, for a dynamic range of 501.

\begin{figure}
\epsfysize=12cm \epsfbox{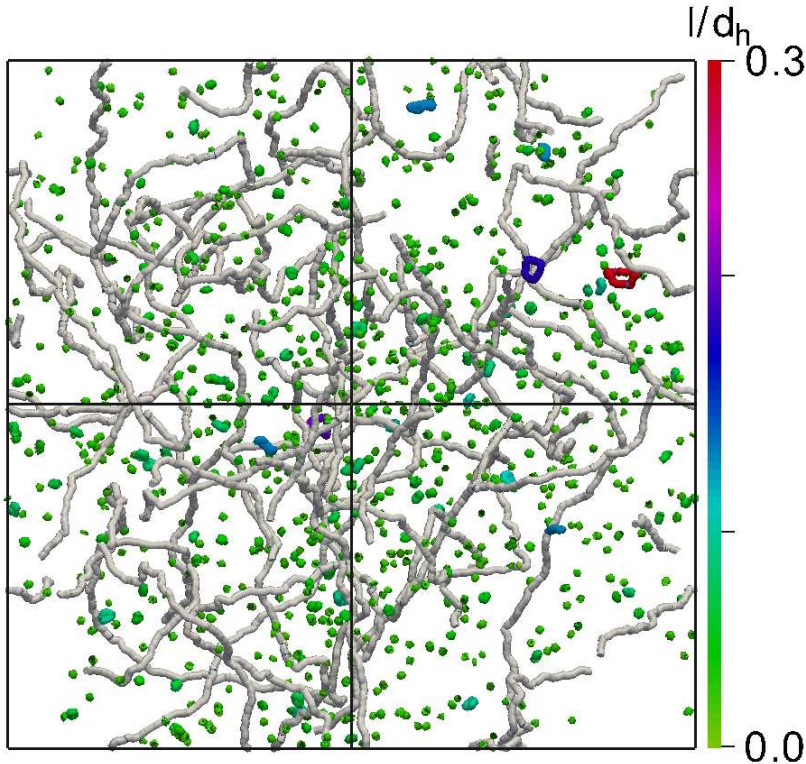}
\caption{A picture of the string network at time 500
  in a flat-space simulation of size 500.  Long strings (here, any
  loop longer than the horizon size) are shown in light gray and loops
  are colored (shaded) according to their length, defined as total
  energy divided by $\mu$.  Non-self-intersecting loops smaller than
  length 10 have been removed.  The edges of the simulation volume are
  shown as black straight lines.  The rhombic dodecahedron is seen
  from one of the order-4 vertices, so that its projection looks
  square.  The side of the square is the horizon distance, 500 units.
  The largest loop, shown in red (dark gray) on the far right
  above the centerline has length about 148.  It appears much smaller
  because it is wrapped into a closed loop, because it has depth that
  cannot be seen, because it is wiggly, and because its length
  includes its kinetic energy.  Loops of this size are rarely seen in
  the loop production function, so this loop will probably fragment or
  rejoin the long string network.}
\label{snapshot}
\end{figure}

These dynamic ranges are much larger than those of any previous
simulation.  The largest simulation before this was done by our group
\cite{OV-06} and claimed dynamic range 60 in the matter era and 120 in
the radiation era.  However, that claim was based on an
artificial setting of the initial clock to 2.0 and 1.0 respectively.
The initial conditions in those simulations were more appropriate for
initial times of order 4.5 and 6.0, giving effective dynamic ranges 27
and 20.  Ref.~\cite{Ringeval-1} used dynamic range 8 in matter and 17
in radiation, and all other previous simulations were smaller.
A snapshot of a flat-space simulation is shown in Fig.~\ref{snapshot}.

\section{Scaling network dynamics}

Suppose that the cosmic string network does evolve into a scaling
regime.  Then we hope that simulations will show an approach to this
regime.  What should we look for?

\subsection{String density}

First, let us consider scaling of the long string energy density, which is the
easiest to find.  Let $d$ denote the average interstring distance of
long strings, defined by
\beq
\label{d-def}
d = \sqrt{\frac{\mu}{\rho_\infty}},
\eeq
where $\mu$ is the tension of the string and
$\rho_\infty$ is the energy density of the long string network.
In a scaling regime we expect this distance to be proportional to the
horizon distance $d_h = a \eta$, so that
\begin{equation}
\gamma = \frac{d}{d_h}
\end{equation}
is a constant.

\subsection{Loops}

Now let us turn to the distribution of loops.  We characterize loops
by their energy $\mu l$, and we call $l$ the length.  Some groups have
concentrated on the the loop density distribution, which we will
denote here as\footnote{But note that \cite{VOV-06,OV-06}
    used $n(l,t)$ to denote the production density, which we call
    $f(l, t)$ here.}
\beq
n(l,t) = \frac{dN}{dl\,dV}
\eeq
so that $n(l,t) dl$ is number of loops in the network at time $t$
whose lengths lie between $l$ and $l + dl$.  Others
concentrate on the density of loop production, which we will denote
\beq
f(l,t) = \frac{dN}{dt\,dl\,dV}
\eeq
so that $f(l,t)\,dt\,dl$ is the number of loops produced per unit volume
with $l \in [l,l + dl]$ at times $t \in [t,t+dt]$.

To put these functions in terms of scaling variables, we trade the
loop length $l$ for a length in scaling units, $x=l/d_h$.  Similarly, we
will use a scaling length interval $dx$ and a scaling spatial volume $d_h^3$
or spacetime volume $d_h^4$.  Thus we define the dimensionless functions
\beq
n(x) = d_h^4 n(l,t)
\eeq
and
\beq
f(x) = d_h^5 f(l,t).
\eeq
We write the left-hand sides as functions of $x$, since in a scaling
regime these functions will depend only on $x$, not on time.

\subsection{Energy balance}

The production of loops and scaling of long strings are related.  If
we consider a long piece of string whose length is $l$, the expansion
of the universe will stretch and redshift it according to the relation
\cite{Kibble-85}
\beq
\label{longldot-eq}
\frac{dl}{dt}= H l \left(1 - 2 \left<v^2_\infty\right>\right)
\eeq
where $\left<v^2_\infty\right>$ is the average squared velocity of the
long strings.  The energy density in long strings then obeys the
Boltzmann equation
\beq
\label{long-rho-dot-eq}
\frac{d\rho_\infty}{dt} = -2 H \left(1 + \left<v^2_\infty\right>
\right)\rho_\infty - \mu \int_0^\infty l f(l,t)\,dl.
\eeq
Now consider a cosmology with scale factor $a \sim t^\nu$, so that
$\nu$ is equal to 0 in flat spacetime, $1/2$ for radiation domination, and $2/3$
for matter domination.  The horizon distance is then $t/(1 - \nu)$.
In a scaling regime $\rho_\infty = \mu/(\gamma^2 d_h^2)$, so we can
rewrite Eq.~(\ref{long-rho-dot-eq}) as
\beq
\int_0^\infty l f(l,t)\,dl = \frac{2}{\gamma^2 d^3_h}
\left(1 - \frac{\nu}{1-\nu}\langle v^2_\infty\rangle\right).
\label{lg-norm-eq}
\eeq
In scaling variables, Eq.~(\ref{lg-norm-eq}) becomes
\beq
\int_0^\infty x f(x)\,dx = 
\frac{2}{\gamma^2}
\left(1 - \frac{\nu}{1-\nu}\langle v^2_\infty\rangle\right).
\label{xf-norm-eq}
\eeq

A consequence of Eq.~(\ref{xf-norm-eq}) is that any scaling loop
production function $f(x)$ must be normalizable, in the sense that the
left-hand side of Eq.~(\ref{xf-norm-eq}) converges.

Below, we will measure $\gamma$ and $\langle v^2_\infty\rangle$
in our simulations and use them to predict the left-hand side of
Eq.~(\ref{xf-norm-eq}), as a check on the consistency of our results.

\subsection{Integrated loop production}
\label{analytic-loop-distribution-sec}

The loop density distribution arises from all prior production of
loops.  We will ignore gravitational damping, which is not included in
our simulation.  We also, at first, ignore the decrease in the energy of
loops from redshifting of their center-of-mass velocities.  In this
approximation, subhorizon loops retain their physical length $l$, and the
only change in the distribution of loops in a comoving volume is due
to the production of new loops from the long string network.  Thus

\begin{equation}
\label{boltzmann-eq}
\frac{d}{dt}\left[a^3(t)n(l,t)\right] = a^3(t)f(l,t).
\end{equation}
Accordingly we can find $n(l,t)$ from $f(l,t)$ by integration,
\beq\label{ng-eq}
n(l,t) = \frac{1}{a^3(t)}\int_0^t a^3(t') f(l,t')\,dt'
= \int_0^t \left(\frac{t'}{t}\right)^{3\nu}f(l,t')\,dt',
\eeq
since $a(t) \sim t^\nu$.

We would like to write this in terms of the scaling functions $n(x) =
d_h^4(t)n(l,t)$ and $f(x) = d_h^5(t)f(l,t)$.  Let us change the
integration variable from $t'$, the creation time of the loop, to $x'
= l/d_h(t')$, the scaling length that it had when it formed.  We similarly
exchange $t'/t$ for $x/x'$ to get
\begin{equation}\label{nx-eq}
n(x) = (1-\nu)x^{3\nu-4}\int_x^\infty x'^{\,3-3\nu} f(x')\,dx'.
\end{equation}
Since $\nu$ is at most $2/3$, $3-3\nu$ is at least $1$.  Since
$f(x)$ must be normalizable (in the sense of Eq.~(\ref{xf-norm-eq})), the integral
on the right hand side of Eq.~(\ref{nx-eq}) converges even if $x$
is taken to $0$.  Thus for loops sufficiently below the horizon size ($x
\ll 1$), Eq.~(\ref{nx-eq}) is insensitive to the lower limit, and we
get a power-law prediction \cite{Alex-book} for the loop density
distribution,
\begin{equation}
\label{N-function}
n(x) = \omega_\nu x^{3\nu-4},
\end{equation}
with
\begin{equation}
\label{nu-eq}
\omega_\nu = (1-\nu)\int_{0}^\infty x^{3-3\nu} f(x)\,dx .
\end{equation}

Now let us consider the effect of the center-of-mass velocity of
emitted loops on Eq.~(\ref{nx-eq}).  We will continue to neglect
gravitational damping.  As we will show below, typical loop velocities
increase with decreasing $x$, and tiny loops are usually emitted with
quite large boosts.  For simplicity, we will assume that the dominant
contribution to Eq.~(\ref{nx-eq}) comes from a single $x_1$
corresponding to a typical loop center-of-mass velocity $v_1$.  For
$x$ near $x_1$, the effects are somewhat complicated \cite{velocity-correction}, and we will not
attempt to model them here, but for $x\ll x_1$ matters are simpler.
In this case, the most important loops have been redshifted
essentially to rest, and thus their energy has decreased by a factor
$\gamma_1 = \left(1-v_1^2\right)^{-1/2}$.

Thus we can update our calculation by taking loops with length $l$ to
have been formed with length $l'=\gamma_1 l$.  Equation~(\ref{ng-eq})
becomes
\beq\label{ngv-eq}
n(l,t) = \gamma_1\int_0^t \frac{t'^{\,3\nu}}{t^{3\nu}}f(l',t')\,dt'.
\eeq
The factor of $\gamma_1$  arises because $n(l,t)$ is the loop
density per unit $l$, while $f(l',t')$ is the loop production density
per unit $l'$ and $dl'/dl = \gamma_1$.

The relationship between $x'$ and $t'$ is now $x'=l'/d_h(t') =
\gamma_1 l/d_h(t')$ so $x/x' = t'/(\gamma_1t)$, and we find once again
that for $x \ll x_1$ we have  Eq.~(\ref{N-function}), but with
\begin{equation}
\label{nu1-eq}
\omega_\nu = (1- \nu)\gamma_1^{3\nu-3}\int_{0}^\infty x^{3-3\nu} f(x)\,dx .
\end{equation}

Equation~(\ref{N-function}) tells us that the scaling loop density
distribution $n(x)$ has a universal form (for small $x$) that does
not depend on the shape of $f(x)$, except in an overall factor.  The
only way that a simulation could find a function $n(l,t)$ whose $l\ll
t$ behavior does not have the form $d_h^{-4} n(x)$ with $n(x)$ given by
Eq.~(\ref{N-function}) is for the loop production in that simulation
to not be in a scaling regime.

This conclusion applies even though real cosmic strings, unlike those
in usual simulations, would be affected by gravitational damping.  It
does not make sense to claim that simulations indicate a
nonnormalizable $f(x)$, but that $f(x)$ will be cut off for low $x$ by
gravitational damping.  If simulations indicate a scaling function
$f(x)$, then if one did a very long simulation, one should observe
$f(x)$ approaching that scaling form.  In order to approach a
nonnormalizable $f(x)$, the fraction of the energy of the string
network emitted into loops in a Hubble time in the simulation would
have to grow without bound.  Such a situation could never be found in
a simulation that conserved energy, so such a result does not make sense.

The same argument applies to analytic claims about loop production
functions.  It is not reasonable to predict a nonnormalizable $f(x)$
on the basis of analytic reasoning that ignores gravitational back
reaction, and then claim that such a thing is made acceptable by the
inclusion of gravity.  What could such analysis predict about strings
with infinitesimal $G\mu$, so that gravity could be ignored?  Some
consistent answer should emerge, which means that the nonnormalizable
growth of $f(x)$ would have to be cut off at some small $x$.  One must
then analyze whether this cutoff is at a smaller or larger scale than
that produced by gravitational damping, which would depend
on the value of $G\mu$ under consideration.


\section{Results: Long string density}

We now present results from our simulations, beginning with the
scaling of the long string density.  Long strings are commonly defined
as those whose energy is above some threshold.  Here, however, we
distinguish the long strings from the loops based on interactions.  If
a loop undergoes 3 oscillations\footnote{This is the minimum number of
  oscillations before our algorithm can confirm that the trajectory is
  non-self-intersecting.  Changing this to a larger number does not
  significantly affect the results.}  without intersecting itself or
another string, we retroactively consider it to be a loop beginning at
the time when it was formed.  Everything else belongs to the long
string network.

\begin{figure}
\epsfysize=10cm \epsfbox{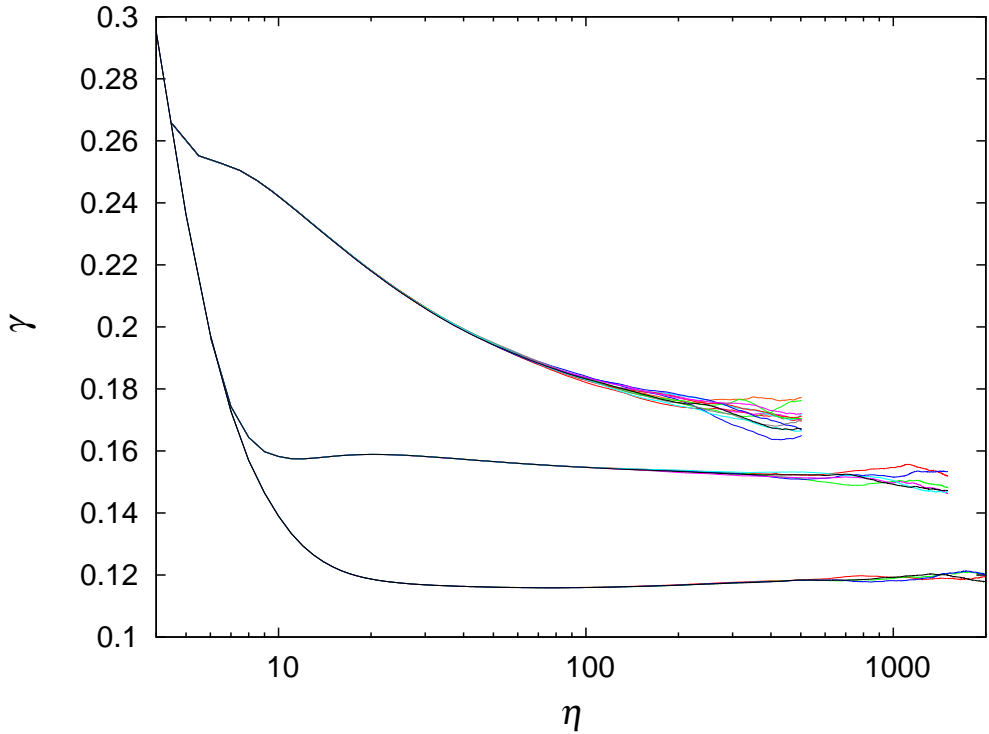}
\caption{The ratio of the interstring distance to the horizon size.
  The top group of lines is for the matter era, the middle group the
  radiation era, and the bottom group flat spacetime.}
\label{isd}
\end{figure}

With this definition, we compute the parameter $\gamma = d/d_h$, with
$d$ given by Eq.~(\ref{d-def}).  For a scaling network we expect
$\gamma$ to achieve a constant value.
In Fig.~\ref{isd}
we plot $\gamma$ as a function of conformal time $\eta$.  The top
group of lines shows 13 runs of size 500 in the matter era, the middle
group 6 runs of size 1500 in the radiation era, and the bottom group 4
runs of size 2000 in flat spacetime.  In all cases, the scaling of
long strings is well established early in the simulation.  We note
that in the flat case, the scaling results entirely from loop
production, as there is no Hubble friction present.

Nevertheless, as seen in this plot against logarithmic time, the
interstring distance is slowly changing, even at end of the
simulations, so the final value cannot be precisely determined.  At
late times, there is also quite a lot of noise because of fluctuations
in the actual long loops produced.

We average the interstring distance over all runs in each era at the
end of the simulation, and show the results in Table \ref{tablegamma},
\begin{table}
\abovecaptionskip 10pt          
\begin{tabular}{|c|c|c|c|c|}
\hline
First author \& Ref. & \hspace{5pt}Flat\hspace{5pt} & Radiation & Matter\\
\hline
Albrecht \cite{Albrecht-Turok-2} & & 0.07 & 0.12 \\
Bennett \cite{Bennett-Bouchet-2} & & 0.14 & 0.18 \\
Allen \cite{Allen-Shellard} & & 0.13 &\\
Vanchurin \cite{VOV-05} & 0.096 & & \\
Ringeval \cite{Ringeval-1}  && 0.162 & 0.188\\
Martins \cite{Martins-Shellard-2} & 0.23 & 0.13 & 0.20\\
Bevis \cite{Bevis-et-al} & & 0.255 & 0.285\\
This paper & 0.12 & 0.15& 0.17\\
\hline
\end{tabular}
\caption{Values of $\gamma$, the ratio of the interstring distance to
  the horizon size, from present and past simulations.  Ref.~\cite{Bevis-et-al}
  used lattice field theory and included all string in $\rho_\infty$.
  All other simulations used Nambu-Goto, and all but the present paper
  included loops larger than some scaling threshold in $\rho_\infty$.
  Here we include all loops which will fragment (or rejoin) in
  $\rho_\infty$.   This is the correct definition for the calculation
  of energy conservation in \S\ref{sec:energy-balance} below.
  Including only loops larger than the horizon size would increase
  $\gamma$ by no more than 3\%.}
\label{tablegamma}
\end{table}
in comparison with previous simulations.  In general we see good
agreement with other recent results. The exception is the field theory
simulations \cite{Bevis-et-al} which do not seem to agree with any
Nambu-Goto simulation, and the flat space simulations of
Ref.~\cite{Martins-Shellard-2}, which gave an interstring distance
twice as large as ours.

The agreement with other simulations whose dynamic range was
significantly smaller than ours supports the conclusion that the long
string component of the network finds its way to a scaling solution on
a relatively short time scale, even if the other properties of the
network have not relaxed to their scaling solutions yet.  We will
comment on this effect again when we discuss the scaling of loops.


\section{Results: Loop production}

In this section we study the loop production function $f(x)$ and the
loop distribution function $n(x)$, as defined above.  We consider a
loop to be produced when it first enters a non-intersecting
trajectory.  Loops that fragment or that rejoin the long string
network are not counted.

The distribution $f(x)\,dx$ gives the spectrum of loop production,
which is not necessarily integrable. Instead we seek the
distribution of power going into loops $x f(x)\,dx$. Due to the wide range of
loop sizes covered by our simulations, it is illuminating to plot this
power density in logarithmic units: $x^2 f(x)\,d\ln x$. In order to
interpret the area under the curves as the total power produced in
loops, we plot $x^2 f(x)$ on linear-log axes in the figures below.

\subsection{Matter era}

We have performed thirteen simulations on a comoving box of $500$
initial correlation lengths each, in a matter era background. The
initial conformal time was set to $\eta_i = 4.5$, and we ran until
$\eta_f = 504.5$. Following the algorithm described above, we removed
the non self-intersecting loops from the simulation, keeping record of
their size and time of production.  Using this information, we can
reconstruct the loop distribution $n(x)$ and the loop production
function $f(x)$ as time progresses to determine if either approaches a
scaling regime.

\begin{figure}
\centering\leavevmode
\epsfysize=11.5cm \epsfbox{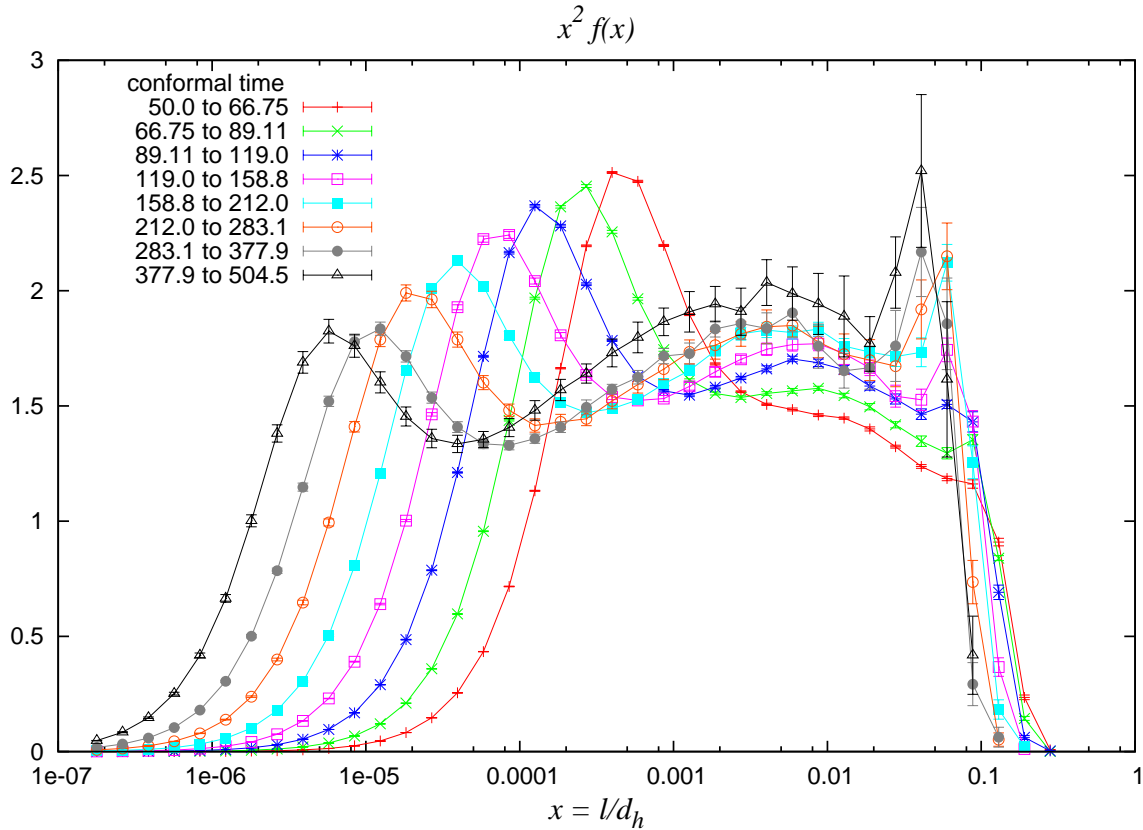}
\caption{The spectrum of loop production power $x^2 f(x)$ in the
  matter era.}
\label{loops-matter}
\end{figure}

We first turn our attention to the loop production function.  We plot
in Fig.~\ref{loops-matter} the data for $f(x)$.
The error bars show one standard deviation from the thirteen successive runs.
The form of the curves in Fig.~\ref{loops-matter} is similar to that
of \cite{OV-06}, but because our simulation is larger than that of \cite{OV-06}
we can follow the evolution of $f(x)$ to much later times.

Notice these curves have a slightly increasing area toward later times.
This suggests the interstring distance $d$ began at a value larger than
its scaling value, and the initially lower power going into loops allowed
it to decrease toward the scaling value.  This is confirmed in
Fig.~\ref{isd}.  Why didn't we choose our initial time so that the
initial $\gamma=d/d_h$ was equal to the scaling value?  The reason
is that we found it more important to have the the height of the small
loop production peak start out unchanging in time, so that we could be
sure that the decrease at later times (shown in Fig.~\ref{loops-matter})
was not an artifact of the initial conditions.  One cannot make both
choices simultaneously because the structures on strings in the
initial conditions are not very close to the scaling regime.

\begin{figure}
\centering\leavevmode
\epsfysize=11.5cm \epsfbox{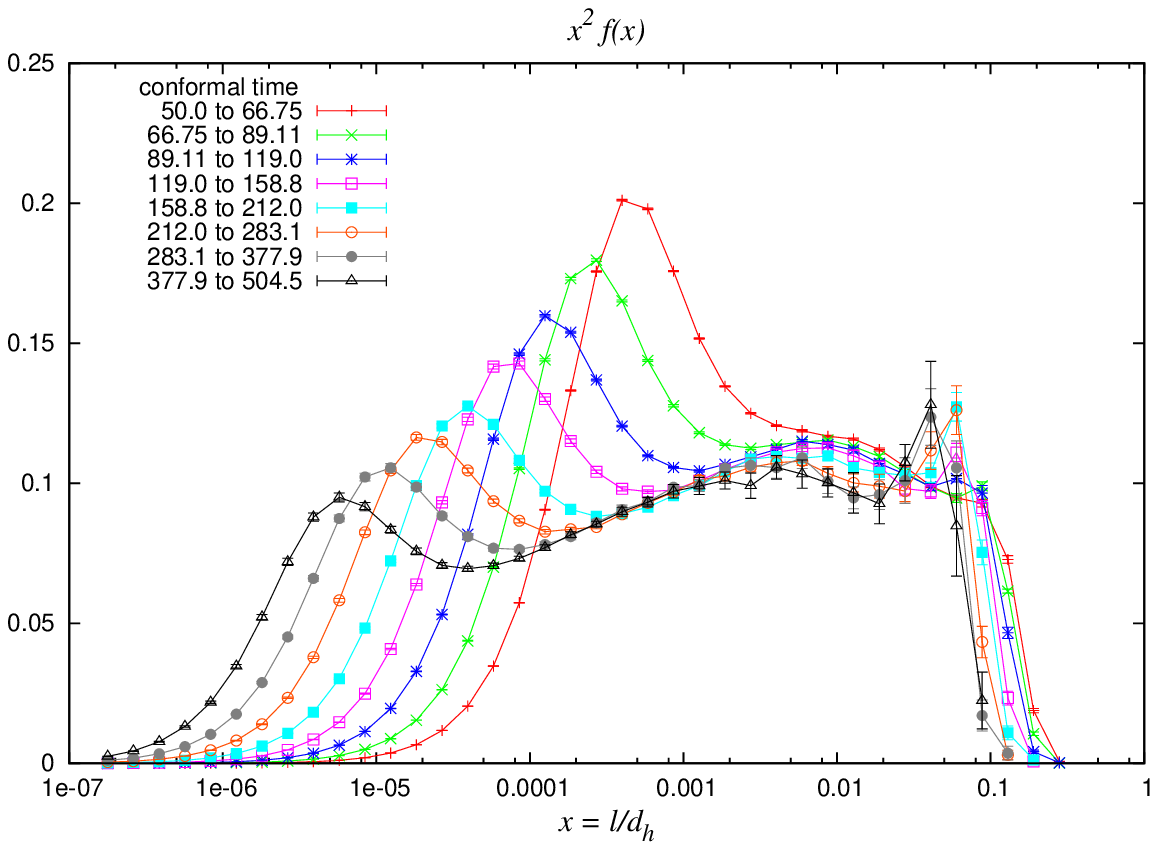}
\caption {Normalized $x^2 f(x)$ during the matter era. 
While perhaps 70\% of the power goes into initial-conditions
loops at conformal time 50.0, less than 25\% does so by conformal time 500.0.}
\label{loops-matter-normalized}
\end{figure}

We can get a better look at the relative shapes of the
curves in Fig.~\ref{loops-matter} by normalizing each curve to unity.
This is done in Fig.~\ref{loops-matter-normalized}, which now
gives the relative flow of power into loops of different sizes.
We see that the normalized $f(x)$ appears to converge more rapidly to
a final scaling form.

We can identify three different peaks in the loop production.  The
first is sharply peaked at $x \approx 0.05$.  The second is centered
approximately an order of magnitude smaller in $x$, and is much wider.
The third peak is clearly moving toward smaller $x$, representing the
transient (non-scaling) population of loops seeded by the initial
conditions.  The non-scaling peak is decreasing in area relative to
the scaling portion of the distribution, and is subdominant after a
conformal time of order 100.  At the farthest reach of our simulation,
the non-scaling power represents of order $25\%$ of the total loop
power, and we expect it to continue to decline in larger simulations.

It is clear from the time dependence of $f(x)$ that the small
scale structure and loop production has not completely reached
scaling, but we consider the late-time behavior of our data to be
evidence that a time-independent loop production
function exists without the aid of an additional smoothing mechanism such as
gravitational radiation.  While it is possible that the non-scaling
third peak will continue to evolve toward smaller $x$ while
maintaining a constant (but certainly subdominant) area, we
do not expect this to be the case.

The other two peaks appear to scale: they do not move toward smaller
$x$ at later times.  We do not know why there are two such peaks, but
one speculation is that one peak represents loops which resulted by
chance from two or more spacelike-separated intercommutations, while
the other represents those that were formed from features on a long
string by a single reconnection.  Another possibility is that the peak
with smaller $x$ represents the results of fragmentation after loops
are formed, while the peak with larger $x$ represents those which
happen to form in non-self-intersecting trajectories.

\begin{figure}
\vspace{-6pt}
\centering\leavevmode
\epsfysize=10cm \epsfbox{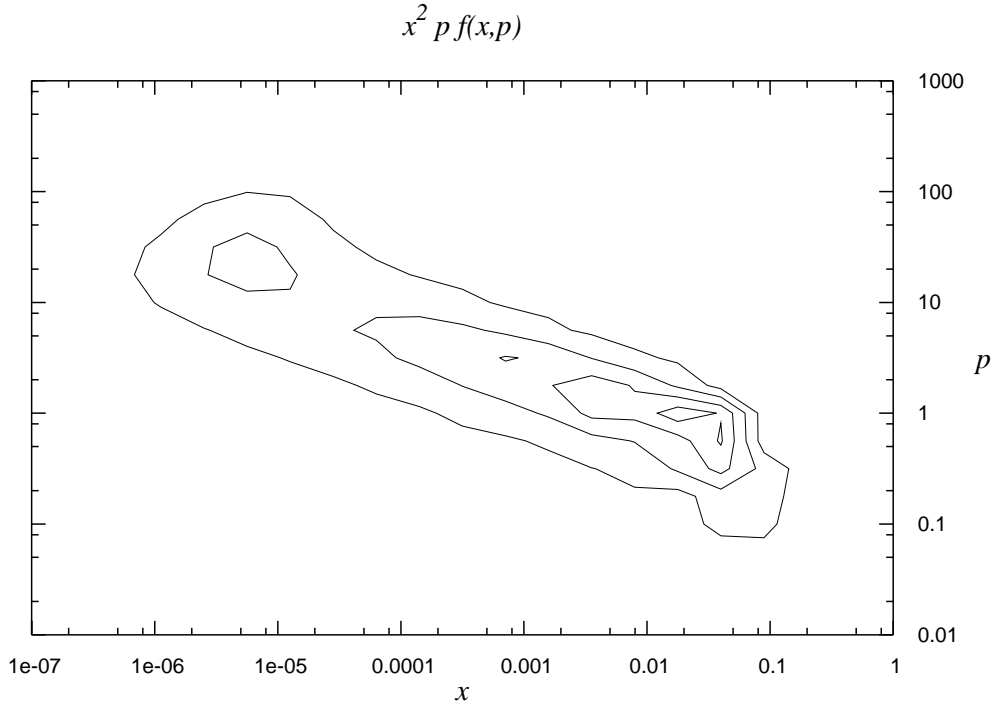}
\caption {Contour plot showing the distribution of loop power in the matter
  era, for conformal times from 283.1 to 504.5.  Here $p = v/\sqrt{1-v^2}$.}
\label{matter-speeds}
\end{figure}

To study the center-of-mass velocities of loops, we define $p =
v/\sqrt{1-v^2}$, the loop momentum per unit rest mass, and let
$f(x,p)\,dx\,dp$ be the scaling production rate of loops with $x \in [x,x + dx]$ and $p
\in [p,p+dp]$.  To show $x f(x,p)\,dx\,dp$ we plot $x^2 p f(x,p)$ on
logarithmic axes in Fig.~\ref{matter-speeds}.  As
expected, small loops at late times are mostly formed with
ultrarelativistic speeds.

\begin{figure}
\centering\leavevmode
\epsfysize=11.5cm \epsfbox{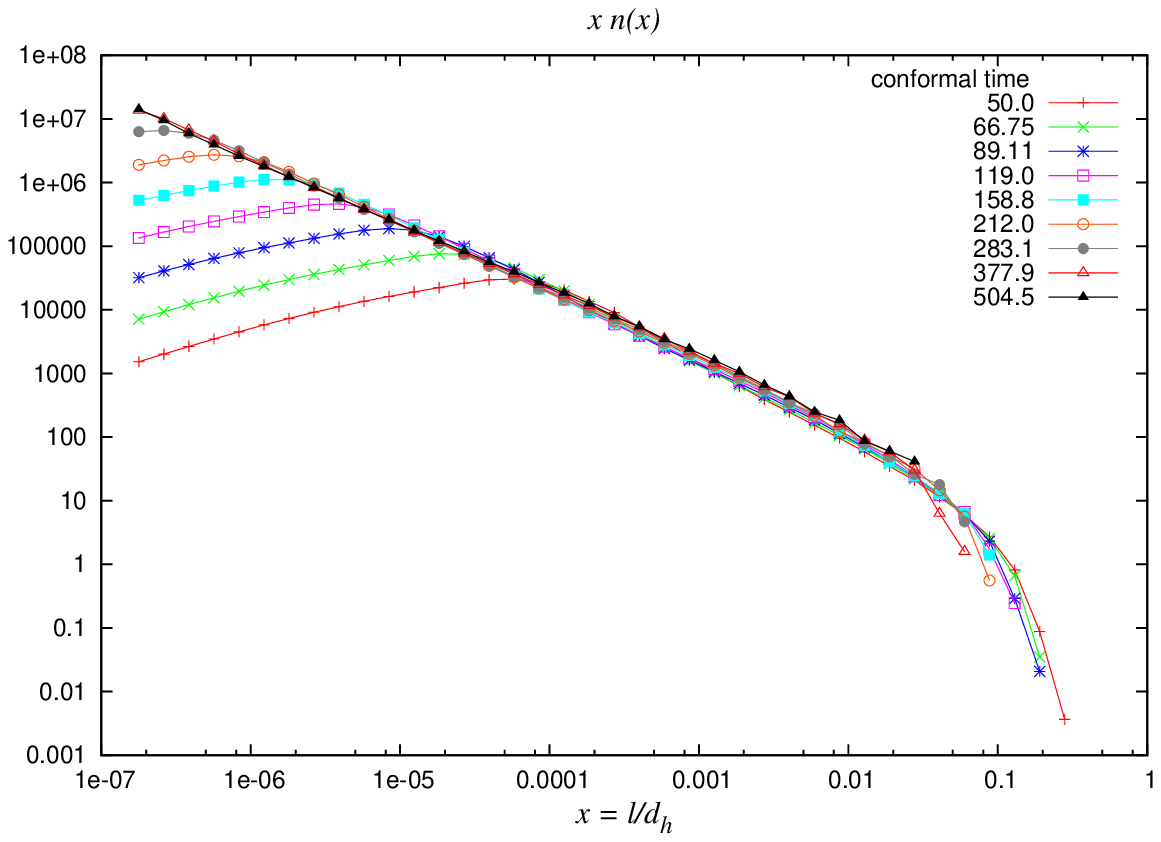}
\caption{The number density distribution of loops $x\,n(x)$
  during the matter era.}
\label{matter-n}
\end{figure}

We now plot in Fig.~\ref{matter-n} the number density distribution
$n(x)\,dx$, which is represented by $x\,n(x)$ on logarithmic axes.  We
see that the solution is in good agreement with the results of
Eq.~(\ref{N-function}), which for the matter era predict a power law
behavior of the form $x^{-1}$. This is different from the result
obtained in \cite{Ringeval-1}, where the power law was found to be of
the form $x^{-1.4}$. As we showed in
\S\ref{analytic-loop-distribution-sec} above, the scaling form of
$x\,n(x)$ could not go as $x^{-1.4}$ throughout the range of $x$ where
loops are produced.  Either there must be significant loop production
in the simulations of \cite{Ringeval-1} at scales below where the
$x^{-1.4}$ fit applies or their loop distribution must not have reached
its final scaling form.

\subsection{Radiation era}

\begin{figure}
\centering\leavevmode
\epsfysize=11.5cm \epsfbox{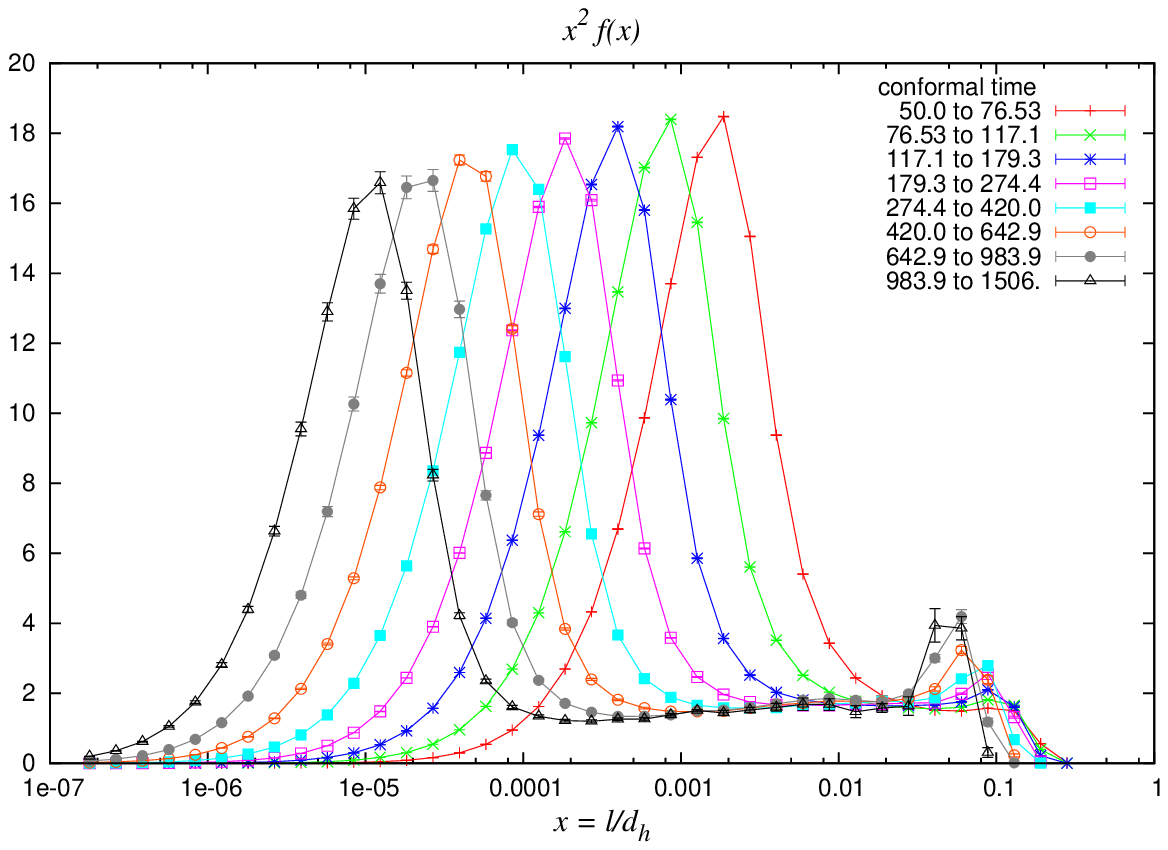}
\caption {The loop production power $x^2 f(x)$ in the radiation era.}
\label{loops-radiation}
\end{figure} 
Fig.~\ref{loops-radiation} shows less dramatic, but similar, scaling
behavior in the radiation era, where we performed six simulations of
size $L = 1500$ comoving initial correlation lengths.  The results are
again qualitatively similar to those of \cite{OV-06}.  Notice that the
total power going into loops is much higher for radiation
domination than for matter.

The approach to loop scaling is much slower (in conformal time) than
occurs in our simulations of the matter era.  Nevertheless, it appears
that a large but shrinking population of transient loops at the
initial conditions scale is making way for a scaling population
sharply peaked at $x = 0.05$, with a nearly flat distribution
extending over many orders of magnitude, perhaps peaked an
order of magnitude smaller in $x$.

We can understand the slower approach to scaling as follows. During
radiation domination, the increase of scale factor with conformal time
is much lower, and so strings are being stretched and thus
smoothed more slowly, causing small-scale structure to persist for longer.
Furthermore, the redshifting of string energy is less efficient than
during the matter era, and therefore the network will need to make
more use of loop production in order to reach a scaling solution, so
the overall production rate is much higher.

We cannot predict the scaling form of $f(x)$ as well as we could
for the matter era, since a dominant fraction of power is in the
transient regime.  But the late-time radiation era power looks similar
to that in the matter era at a much earlier time.  We expect that the
radiation era evolution should be similar to that in the matter era,
with the scaling features eventually dominating, and $f(x)$ eventually
approaching a time-independent form, but  the approach to
scaling will last much longer.

\begin{figure}
\vspace{-6pt}
\centering\leavevmode
\epsfysize=10cm \epsfbox{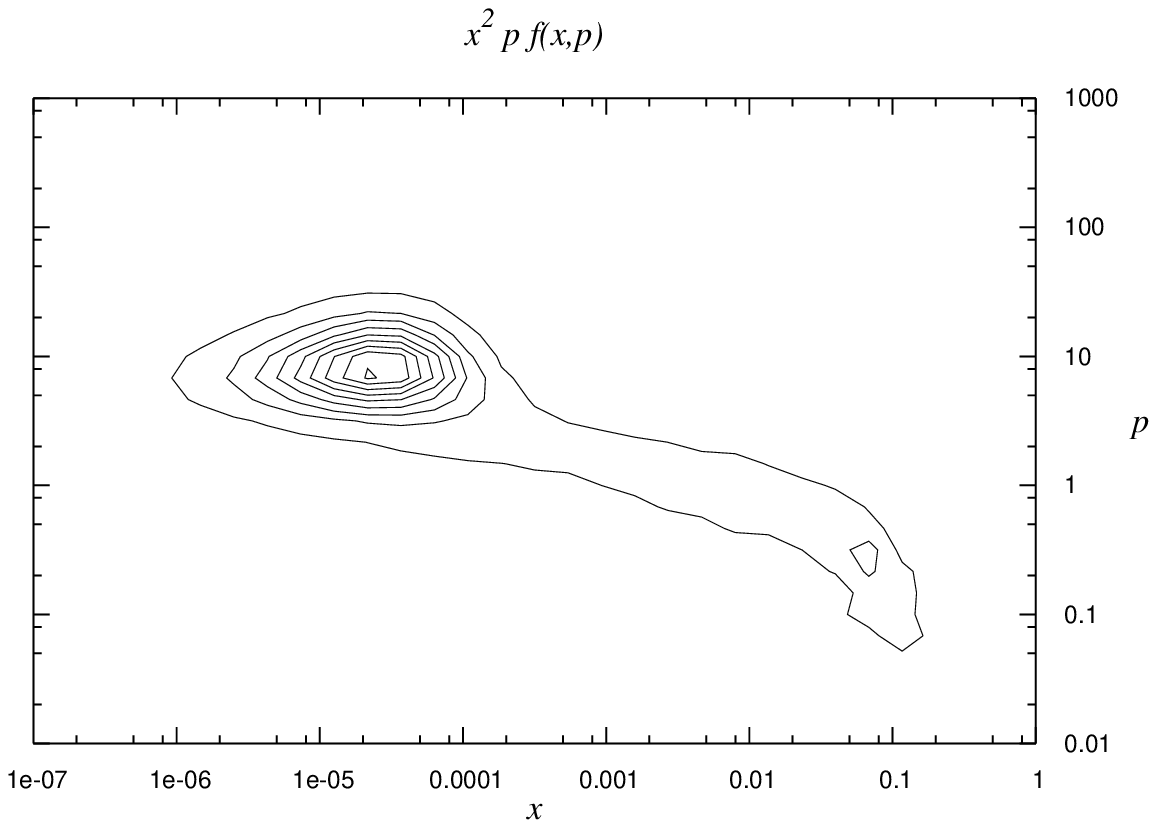}
\caption{Contour plot showing the distribution of loop power in the radiation
  era, for conformal times from 475 to 1006. Here $p = v/\sqrt{1 - v^2}$.}
\label{radiation-speeds}
\end{figure}

The velocity dependence of the loop production function can be seen in
Fig.~\ref{radiation-speeds}.

\begin{figure}
\centering\leavevmode
\epsfysize=11.5cm \epsfbox{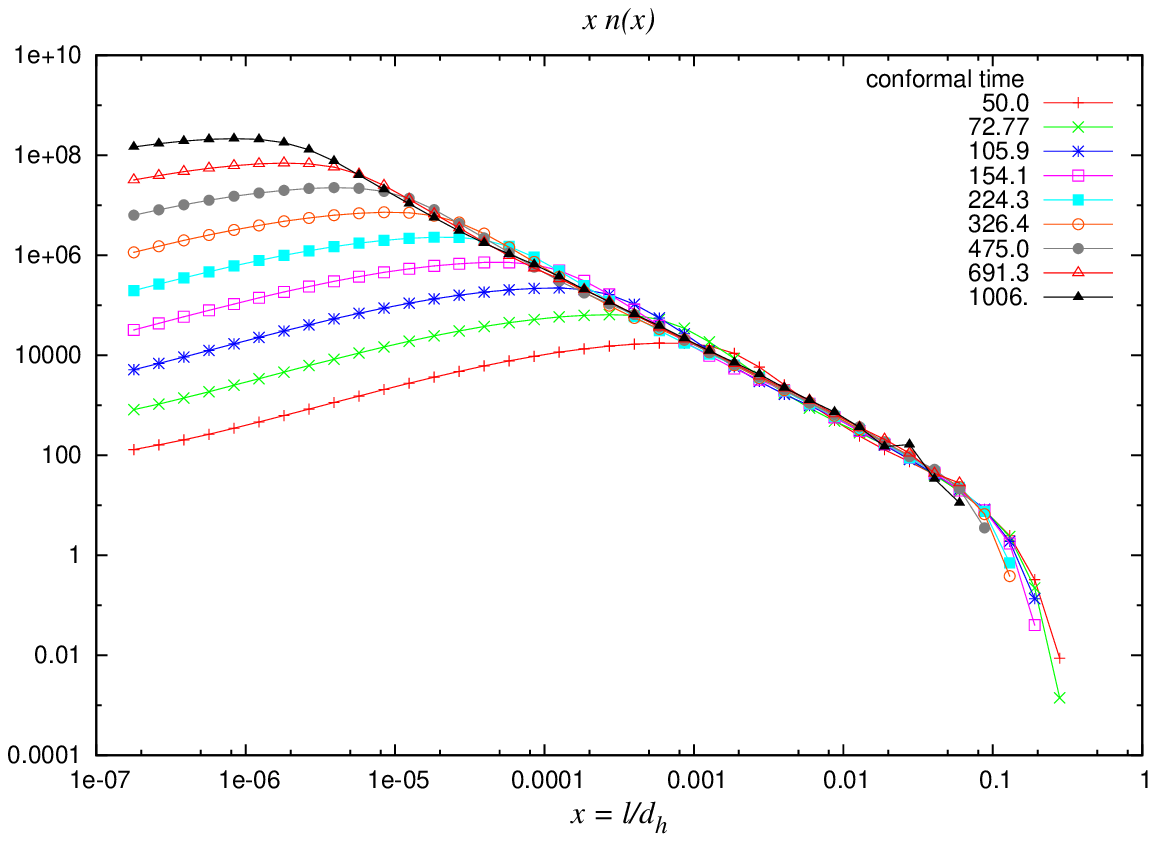}
\caption{The number density distribution of loops $x\,n(x)$
  during the radiation era.}
\label{radiation-n}
\end{figure}
We show in Fig.~\ref{radiation-n} our dataset $n(x)$ binned
logarithmically from the simulation. The data obtained this way agrees
quite well with the analytical results presented earlier for the
radiation era, where $x\,n(x) = \omega_\frac{1}{2}x^{-3/2}$.  This
should not be misinterpreted as true scaling, since we know that a
dominant portion of the power in loops is in a transient, not a
scaling population.

Roughly speaking, there are three possibilities which can occur at
very late times (assuming $\gamma_r$ is static).  The non-scaling peak
could smear itself out as it continues to move toward smaller $x$
while creating an extremely broad plateau.  It could conceivably move
off the plateau and continue indefinitely toward smaller $x$ while
maintaining a constant area.  It could also decline while the peak at
$x = 0.05$ grows, and not leave a tremendous plateau.  We conjecture
the last to be the case.  It should be pointed out that the first
two cases are essentially identical in terms of the prediction for
$\omega_\frac{1}{2}$, but that third case will cause a relative
increase by up to an order of magnitude.  It is this uncertainty which
prevents us from claiming that the loop density shown in
Fig.~\ref{radiation-n} is in the final scaling regime.  This
illustrates why the spectrum of loop power $f(x)$ is a more precise
indicator of scaling than $n(x)$, the spectrum of existing loops.

\subsection{Flat space}

\begin{figure}
\centering\leavevmode
\epsfysize=11.5cm \epsfbox{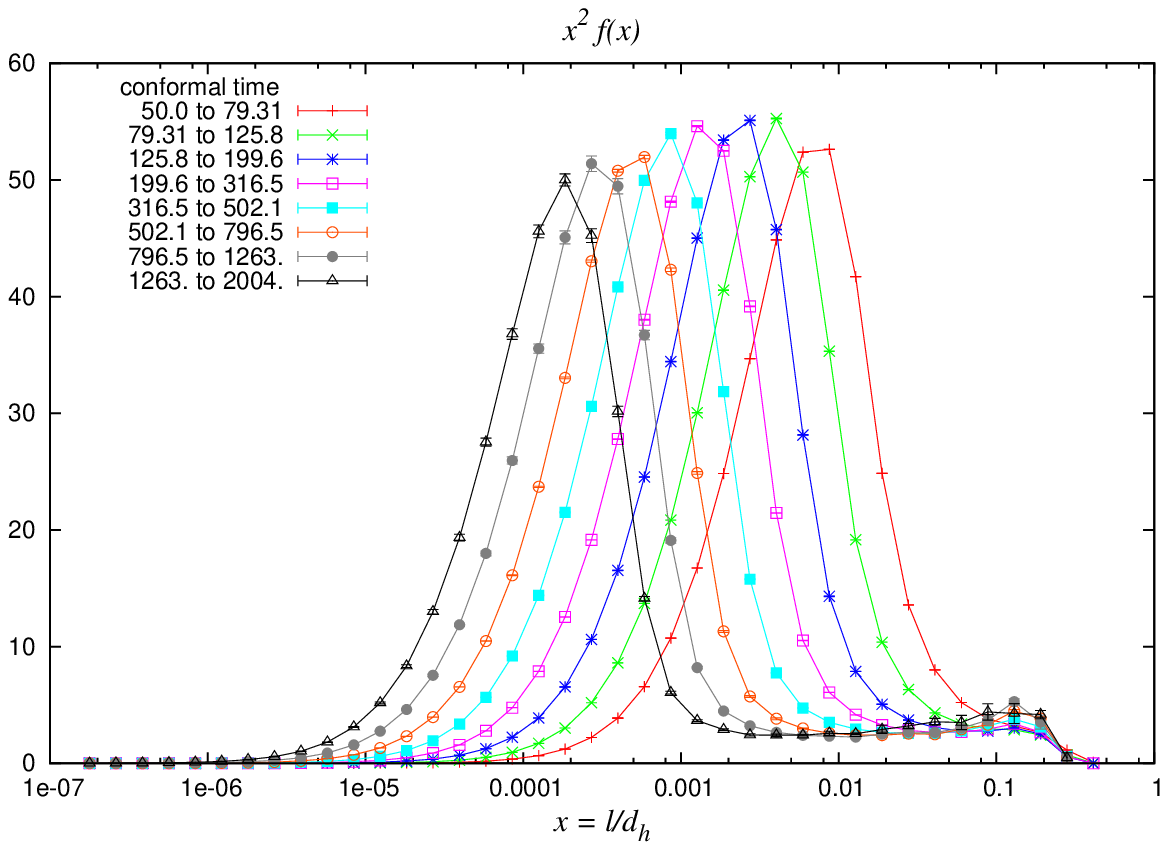}
\caption{The loop production power $x^2 f(x)$ in flat spacetime.}
\label{loops-flat}
\end{figure}
We present in Fig.~\ref{loops-flat} the results of our four
simulations performed in a box of size 2000.  Because redshifting is
nonexistent, all smoothing of the strings comes from loop
production. This massive loop production clearly smooths the strings
enough for a scaling peak at $x \approx 0.1$ to appear, although it
remains small even until conformal time 2000.

The results in Fig.~\ref{loops-flat} are qualitatively similar to
those of \cite{VOV-06}, but the ratio of the scaling to the
non-scaling peak height in \cite{VOV-06} was significantly larger.
There are several differences between the present simulation and that
of \cite{VOV-06}, which make a direct comparison difficult.
Ref.~\cite{VOV-06} achieved box size 800 by a technique of successive
doublings of the box size \cite{VOV-05}, whereas we simulate a box of
size 2000 directly.  The box-doubling technique required
intercommutation probability $P=0.5$, whereas we use $P=1$.  We run
only for an interval of conformal time equal to the box size $L$,
whereas \cite{VOV-06} ran significantly beyond conformal time 800.
Finally, \cite{VOV-06} used somewhat smoother initial conditions in
which a point was interpolated between each pair of successive
Vachaspati-Vilenkin faces, whereas we draw a straight line between
those points.  Smoother initial conditions may have led to a lower
peak at the initial-condition scale.  Note also that \cite{VOV-06,OV-06}
show $x^2 f(x)$ on a logarithmic scale, whereas the vertical axis in
Fig.~\ref{loops-flat} is linear.

Perhaps the most important features of Fig.~\ref{loops-flat} are the
turnaround of the transient peak and the existence of a scaling peak.
After a time $\eta \approx 150$, the amplitude of the transient peak
begins to decrease.  Following the arguments presented above, we again
expect that this decrease will continue, leading to a loop production
function consisting of a time-independent distribution peaked at $x
\approx 0.1$.

\subsection{Energy balance check}\label{sec:energy-balance}

Having measured values for $\gamma$ and $\left<v^2_\infty\right>$
directly from the simulations we can now use Eq.~(\ref{xf-norm-eq}) to
predict the total power emitted by the long string network in the
form of loops, ${\cal P}_{\text{Prediction}}$, and compare it with the direct computation of
this power from the integral of the loop production function, namely,
\beq
{\cal P}_{\text{Simulation}}= \int_0^\infty x f(x)\,dx.
\eeq
\begin{table}
\abovecaptionskip 10pt          
\begin{tabular}{|l|c|c|c|c|}
\hline
 & $~~~~~\gamma~~~~~$ & $~~\left<v^2_\infty\right>~~$ & ${\cal P}_{\text{Prediction}}$ & ${\cal P}_{\text{Simulation}}$\\
\hline
Matter & 0.17 & 0.35 & 21 & 19\\
Radiation & 0.15 & 0.40 & 53 & 51\\
Flat & 0.12 & 0.45 & 139 & 136\\
\hline
\end{tabular}
\caption{Predictions from energy balance.}\label{tableenergybalance}
\end{table}
We show in Table \ref{tableenergybalance} the comparison of these quantities for 
the matter, radiation and flat eras. We see that the agreement between
the predicted result and the numerically found value is within the
statistical noise  in all cases. We consider this is an important check for our
simulations since it serves as a nontrivial test of our code and
algorithms.


\section{Conclusions}
We have developed a new parallel computing technique
\cite{numerical-techniques-paper} that has allowed us to run the
largest cosmic string simulations ever performed, reaching an increase
in dynamic range of roughly an order of magnitude with respect to
previous studies. The results of our simulations indicate a much
slower approach to scaling for loops than for the long string
contribution to the network. This clearly shows the need to push the
dynamic range of the simulation to its largest possible extent in
order to prevent contamination by transient states.

How is it possible for the interstring distance to be quite close to
the final scaling value at early times, while the loop production
function is still very far from scaling?  Surely an important part (in
flat space, the only part) of the loss of energy from the long string
network is due to the production of loops, so if the loop production
mechanism is not in a scaling regime, why does the energy loss rate
scale?

The explanation is presumably that described by Bouchet
\cite{Bouchet-conference}.  The loops are produced by a two-step
process.  The first step is the intercommutation of infinite strings,
which maintains the interstring distance by the usual feedback
mechanism in which an increase in the string network density leads to
more intercommutations.  The second step is the formation of loops
triggered by or made up of the large kinks formed in the
intercommutations.  It controls how the removed string ends up in
loops of different sizes.  Thus the first step can be scaling long
before the second.

Our simulations confirm the results of \cite{VOV-06,OV-06}: There is
an initially dominant transient feature in the loop production
function which very slowly subsides to reveal a significant fraction
of loops produced in a scaling regime.  In the matter era, we find
this population dominates over non-scaling loop production.  Because
this does not occur until extremely late times, we do not believe that
prior claims of having found the final form of loop scaling in
numerical simulations are correct.

In light of our results, we expect that the final loop production
function will eventually be dominated by a broad peak of loop sizes
proportional to the horizon size but ranging downward from about a
twentieth of that size.  We achieve this to good accuracy only in
the matter era, where approximately 75\% of loop power has achieved
scaling by a conformal time of 500.  This will clearly have an impact
on the observational signatures of string networks. We will report on
this in future publications.

\acknowledgments

We would like to thank Vitaly Vanchurin and Alex Vilenkin for helpful
conversations. This work was supported in part by the National Science
Foundation under grant numbers 0855447 and 0903889. These simulations
were performed using a total of 27 years of CPU time on the Tufts
High-performance Computing Research Cluster.  We are also grateful for
access to the Nemo computing facility at the University of
Wisconsin--Milwaukee.  Nemo is funded in part by the National Science
Foundation under grant 0923409.

\appendix


\section{Infinitely difficult cosmic string loops}
We have identified some new features of Nambu-Goto cosmic strings
which are impractical to faithfully simulate.  One of these features
can be called a ``skipping stone."  It occurs in particular cases when
a network contains a loop which to good approximation is piecewise
linear with five or six kinks, and which collides with a long straight
segment of string.  The collision may occur in such a way as to break
off a loop again immediately afterward, one with less energy but
precisely the same shape.

The problem is that the smaller loop will then repeat this process,
skipping off the long string again and again, losing energy each time,
but never changing shape.  Just as a skipping stone leaves behind a
geometric series of ripples, this loop will perform (in the Nambu-Goto
approximation) an infinite number of intercommutations, leaving behind
a geometric series of kinks on the string.

Such a physical process represents a nightmare for a numerical
simulation which has no minimum resolution, since each of these
ripples will be recorded and evolved.

To avoid the tremendous computational resources required to simulate
these rare skipping stones all the way down to the minimum size set by
the floating point resolution, we have intentionally failed to perform
a certain, very small number of intercommutations. These avoided
intercommutations are chosen to only include the collision of a
disjoint loop with another string, and so never prevent the formation
of a loop.  Furthermore, the avoided intercommutation must occur
within a very short distance of a very large number of kinks, i.e.,
only after the stone has skipped at least 50 times do we allow it to
pass through the surface of the water. The fraction of
intercommutations we ignore is less than 0.2\%.

Although these features occur with a varying degree of severity, and
rarely plague small simulations, in a simulation of size 2000 they are
virtually guaranteed to occur at least once with enough alignment so
as to bog down one computer for literally days as it attempts to
evolve through an amount of simulation 4-volume which normally takes a
few seconds.

If cosmic strings exist, we expect that ``infinite repetition''
phenomena such as we discuss here occur in the real universe.  Real
cosmic strings would not have exactly straight segments, but this is
of little consequence.  If an approximation of the ``skipping stone''
phenomenon began, each successive iteration would involve shorter
pieces of the original strings, which would thus be closer to
straight.  Of more importance is that the kinks separating the
straight segments would not be infinitely sharp.  Some kinks appear
anew each cycle, but some remain from the original shape of the loop,
and these would have been smoothed by gravitational back-reaction.
Thus the gravitational damping scale would set a lower bound on the
size of self-similar loops that could be produced.  Since the number
of breakings off and rejoinings grows only logarithmically with the
ratio of the loop size to the curvature radius at the kink, and since
the entire phenomenon is quite rare, we do not expect it to have any
observable consequences.


\begin{thebibliography}{9}


\bibitem{Alex-book}
  A. Vilenkin and E.P.S. Shellard,
  \emph{Cosmic Strings and Other Topological Defects},
Cambridge University Press, Cambridge, England, (1994).

\bibitem{Witten}
  E.~Witten,
  ``Cosmic Superstrings,''
  Phys.\ Lett.\  {\bf B153}, 243 (1985).

\bibitem{Sarangi-Tye}
  S.~Sarangi and S.-H.~H.~Tye,
  ``Cosmic string production towards the end of brane inflation,''
  Phys.\ Lett.\  B {\bf 536}, 185 (2002)

\bibitem{Dvali-Vilenkin}
  G.~Dvali and A.~Vilenkin,
  ``Formation and evolution of cosmic D-strings,''
  JCAP {\bf 0403}, 010 (2004).

\bibitem{Copeland-Myers-Polchinski}
  E.~J.~Copeland, R.~C.~Myers and J.~Polchinski,
  ``Cosmic F- and D-strings,''
  JHEP {\bf 0406}, 013 (2004).


\bibitem{Kaiser}
  N.~Kaiser and A.~Stebbins,
  ``Microwave Anisotropy Due To Cosmic Strings,''
  Nature {\bf 310}, 391 (1984).



\bibitem{gravitational-lenses}
  A.~Vilenkin,
  ``Cosmic Strings As Gravitational Lenses,''
  Astrophys.\ J.\  {\bf 282}, L51 (1984);
  J.~R.~I.~Gott,
  ``Gravitational lensing effects of vacuum strings: Exact solutions,''
  Astrophys.\ J.\  {\bf 288}, 422 (1985);
  A.~Vilenkin, 
  ``Looking For Cosmic Strings," Nature (London), {\bf 322}, 613 (1986);
 B.~Shlaer and S.-H.~H.~Tye,
  ``Cosmic string lensing and closed time-like curves,''
  Phys.\ Rev.\  D {\bf 72}, 043532 (2005).


\bibitem{bounds-from-CMB}

  M.~Wyman, L.~Pogosian and I.~Wasserman,
  ``Bounds on cosmic strings from WMAP and SDSS,''
  Phys.\ Rev.\  D {\bf 72}, 023513 (2005)
  [Erratum-ibid.\  D {\bf 73}, 089905 (2006)];
   A.~A.~Fraisse,
  ``Limits on Defects Formation and Hybrid Inflationary Models with Three-Year
  WMAP Observations,''
  JCAP {\bf 0703}, 008 (2007).
 

\bibitem{bounds-from-CMB-field-theory}

  N.~Bevis, M.~Hindmarsh, M.~Kunz and J.~Urrestilla,
  ``CMB power spectrum contribution from cosmic strings using
  field-evolution simulations of the Abelian Higgs model,''
  Phys.\ Rev.\  D {\bf 75}, 065015 (2007);
  N.~Bevis, M.~Hindmarsh, M.~Kunz and J.~Urrestilla,
  ``Fitting CMB data with cosmic strings and inflation,''
  Phys.\ Rev.\ Lett.\  {\bf 100}, 021301 (2008);
 J.~Urrestilla, N.~Bevis, M.~Hindmarsh, M.~Kunz and A.~R.~Liddle,
  ``Cosmic microwave anisotropies from BPS semilocal strings,''
  JCAP {\bf 0807}, 010 (2008);
  P.~Mukherjee, J.~Urrestilla, M.~Kunz, A.~R.~Liddle, N.~Bevis and M.~Hindmarsh,
  ``Detecting and distinguishing topological defects in future data
  from the CMBPol satellite,''
  Phys.\ Rev.\  D {\bf 83}, 043003 (2011).


\bibitem{Bevis-et-al}
 N.~Bevis, M.~Hindmarsh, M.~Kunz and J.~Urrestilla,
 ``CMB power spectra from cosmic strings: predictions for the Planck
 satellite and beyond,''
  Phys.\ Rev.\  D {\bf 82}, 065004 (2010).



\bibitem{gravitational-radiation}
  A.~Vilenkin,
  ``Gravitational Radiation From Cosmic Strings,''
  Phys.\ Lett.\  B {\bf 107}, 47 (1981);
  T.~Vachaspati and A.~Vilenkin,
  ``Gravitational Radiation From Cosmic Strings,''
  Phys.\ Rev.\  D {\bf 31}, 3052 (1985);
 C.~J.~Burden,
  ``Gravitational Radiation From A Particular Class Of Cosmic Strings,''
  Phys.\ Lett.\  B {\bf 164}, 277 (1985);
 D.~Garfinkle and T.~Vachaspati,
``Radiation from kinky, cuspless cosmic loops,''
  Phys.\ Rev.\  D {\bf 36}, 2229 (1987).


\bibitem{gravitational-radiation-bursts}
  T.~Damour and A.~Vilenkin,
  ``Gravitational wave bursts from cosmic strings,''
  Phys.\ Rev.\ Lett.\  {\bf 85}, 3761 (2000);
  T.~Damour and A.~Vilenkin,
  ``Gravitational wave bursts from cusps and kinks on cosmic strings,''
  Phys.\ Rev.\  D {\bf 64}, 064008 (2001);
  X.~Siemens, J.~Creighton, I.~Maor, S.~Ray Majumder, K.~Cannon and J.~Read,
  ``Gravitational wave bursts from cosmic (super)strings: Quantitative
  analysis and constraints,''
  Phys.\ Rev.\  D {\bf 73}, 105001 (2006);
  C.~J.~Hogan,
  ``Gravitational waves from light cosmic strings: Backgrounds and bursts with
  large loops,''
  Phys.\ Rev.\  D {\bf 74}, 043526 (2006);
 S.~Olmez, V.~Mandic and X.~Siemens,
  ``Gravitational-Wave Stochastic Background from Kinks and Cusps on Cosmic
  Strings,''
  Phys.\ Rev.\  D {\bf 81}, 104028 (2010).



\bibitem{particle-radiation}
  T.~Vachaspati, A.~E.~Everett and A.~Vilenkin,
  ``Radiation From Vacuum Strings And Domain Walls,''
  Phys.\ Rev.\  D {\bf 30}, 2046 (1984);
  R.~H.~Brandenberger,
  ``On the decay of cosmic string loops,''
  Nucl.\ Phys.\  B {\bf 293}, 812 (1987);
  M.~Mohazzab,
  ``Cusp annihilation on ordinary cosmic strings,''
  Int.\ J.\ Mod.\ Phys.\  D {\bf 3}, 493 (1994);
  J.~J.~Blanco-Pillado and K.~D.~Olum,
  ``The form of cosmic string cusps,''
  Phys.\ Rev.\  D {\bf 59}, 063508 (1999);
  K.~D.~Olum and J.~J.~Blanco-Pillado,
  ``Field theory simulation of Abelian-Higgs cosmic string cusps,''
  Phys.\ Rev.\  D {\bf 60}, 023503 (1999).

\bibitem{particle-radiation-from-field-theory}
  G.~Vincent, N.~D.~Antunes and M.~Hindmarsh,
  ``Numerical simulations of string networks in the Abelian-Higgs model,''
  Phys.\ Rev.\ Lett.\  {\bf 80}, 2277 (1998).
 
 

\bibitem{our-field-theory}
K.~D.~Olum and J.~J.~Blanco-Pillado,
  ``Radiation from cosmic string standing waves,''
  Phys.\ Rev.\ Lett.\  {\bf 84}, 4288 (2000).

\bibitem{Moore-field-theory}
  J.~N.~Moore, E.~P.~S.~Shellard and C.~J.~A.~Martins,
  ``On the evolution of Abelian-Higgs string networks,''
  Phys.\ Rev.\  D {\bf 65}, 023503 (2001).

\bibitem{axionic-radiation}
  R.~L.~Davis,
  ``Cosmic Axions from Cosmic Strings,''
  Phys.\ Lett.\  B {\bf 180}, 225 (1986);
 A.~Vilenkin and T.~Vachaspati,
 ``Radiation of Goldstone bosons from cosmic strings,''
  Phys.\ Rev.\  D {\bf 35}, 1138 (1987);
 R.~L.~Davis and E.~P.~S.~Shellard,
  ``Do Axions Need Inflation?,''
  Nucl.\ Phys.\  B {\bf 324}, 167 (1989);
 R.~A.~Battye and E.~P.~S.~Shellard,
  ``Global string radiation,''
  Nucl.\ Phys.\  B {\bf 423}, 260 (1994).



\bibitem{dilaton-radiation}
  T.~Damour and A.~Vilenkin,
  ``Cosmic strings and the string dilaton,''
  Phys.\ Rev.\ Lett.\  {\bf 78}, 2288 (1997);
  E.~Babichev and M.~Kachelriess,
  ``Constraining cosmic superstrings with dilaton emission,''
  Phys.\ Lett.\  B {\bf 614}, 1 (2005);



 \bibitem{light-radiation}
 M.~Srednicki and S.~Theisen,
  ``Nongravitational Decay Of Cosmic Strings,''
  Phys.\ Lett.\  B {\bf 189}, 397 (1987);
 M.~Peloso and L.~Sorbo,
  ``Moduli from cosmic strings,''
  Nucl.\ Phys.\  B {\bf 649}, 88 (2003);
 E.~Sabancilar,
  ``Cosmological Constraints on Strongly Coupled Moduli from Cosmic Strings,''
  Phys.\ Rev.\  D {\bf 81}, 123502 (2010);
  T.~Vachaspati,
  ``Cosmic Rays from Cosmic Strings with Condensates,''
  Phys.\ Rev.\  D {\bf 81}, 043531 (2010);
  D.~A.~Steer and T.~Vachaspati,
  ``Light from Cosmic Strings,''
  arXiv:1012.1998 [hep-th].

\bibitem{CR-from-strings}
  V.~Berezinsky, P.~Blasi and A.~Vilenkin,
  ``Ultra high energy gamma rays as signature of topological defects,''
  Phys.\ Rev.\  D {\bf 58}, 103515 (1998);
  P.~Bhattacharjee and G.~Sigl,
  ``Origin and propagation of extremely high energy cosmic rays,''
  Phys.\ Rept.\  {\bf 327}, 109 (2000).

 
\bibitem{Vilenkin-81}
  A.~Vilenkin,
  ``Cosmological Density Fluctuations Produced by Vacuum Strings,''
  Phys.\ Rev.\ Lett.\  {\bf 46}, 1169-1172 (1981).

\bibitem{Kibble-85}
  T.~W.~B.~Kibble,
  ``Evolution Of A System Of Cosmic Strings,''
  Nucl.\ Phys.\  {\bf B252}, 227 (1985).




\bibitem{Bennett}
  D.~P.~Bennett,
  ``The Evolution Of Cosmic Strings,''
  Phys.\ Rev.\  {\bf D33}, 872 (1986);
  D.~P.~Bennett,
  ``Evolution Of Cosmic Strings. 2.,''
  Phys.\ Rev.\  {\bf D34}, 3592 (1986).
  


\bibitem{Albrecht-Turok-1}
  A.~Albrecht and N.~Turok,
  ``Evolution Of Cosmic Strings,''
  Phys.\ Rev.\ Lett.\  {\bf 54}, 1868 (1985).


\bibitem{Albrecht-Turok-2}
  A.~Albrecht and N.~Turok,
  ``Evolution Of Cosmic String Networks,''
  Phys.\ Rev.\  D {\bf 40}, 973 (1989).



\bibitem{Bennett-Bouchet-1}
  D.~P.~Bennett and F.~R.~Bouchet,
  ``Evidence For A Scaling Solution In Cosmic String Evolution,''
  Phys.\ Rev.\ Lett.\  {\bf 60}, 257 (1988);  D.~P.~Bennett and F.~R.~Bouchet,
  `` Cosmic string evolution,''
  Phys.\ Rev.\ Lett.\  {\bf 63}, 2776 (1989).

\bibitem{Bennett-Bouchet-2}
  D.~P.~Bennett and F.~R.~Bouchet,
  ``High resolution simulations of cosmic string evolution. 1 Network evolution,''
  Phys.\ Rev.\  D {\bf 41}, 2408 (1990).


\bibitem{Allen-Shellard}
  B.~Allen and E.~P.~S.~Shellard,
  ``Cosmic string evolution: A numerical simulation,''
  Phys.\ Rev.\ Lett.\  {\bf 64}, 119 (1990).



\bibitem{Austin-Copeland-Kibble}
  D.~Austin, E.~J.~Copeland and T.~W.~B.~Kibble,
  ``Evolution Of Cosmic String Configurations,''
  Phys.\ Rev.\  D {\bf 48}, 5594 (1993).


\bibitem{Martins-Shellard-1}
  C.~J.~A.~P.~Martins, E.~P.~S.~Shellard,
  ``Quantitative string evolution,''
  Phys.\ Rev.\  {\bf D54}, 2535-2556 (1996).


\bibitem{Polchinski-et-al}
  J.~Polchinski and J.~V.~Rocha,
  ``Analytic Study of Small Scale Structure on Cosmic Strings,''
  Phys.\ Rev.\  D {\bf 74}, 083504 (2006);
  J.~Polchinski and J.~V.~Rocha,
  ``Cosmic string structure at the gravitational radiation scale,''
  Phys.\ Rev.\  D {\bf 75}, 123503 (2007).
  F.~Dubath, J.~Polchinski and J.~V.~Rocha,
  ``Cosmic String Loops, Large and Small,''
  Phys.\ Rev.\  D {\bf 77}, 123528 (2008);


\bibitem{VOV-05}
  V.~Vanchurin, K.~Olum and A.~Vilenkin,
  ``Cosmic string scaling in flat space,''
  Phys.\ Rev.\  D {\bf 72}, 063514 (2005).


\bibitem{Ringeval-1}
  C.~Ringeval, M.~Sakellariadou and F.~Bouchet,
  ``Cosmological evolution of cosmic string loops,''
  JCAP {\bf 0702}, 023 (2007).


\bibitem{Martins-Shellard-2}
  C.~J.~A.~P.~Martins and E.~P.~S.~Shellard,
  ``Fractal properties and small-scale structure of cosmic string networks,''
  Phys.\ Rev.\  D {\bf 73}, 043515 (2006).

\bibitem{VOV-06}
  V.~Vanchurin, K.~D.~Olum and A.~Vilenkin,
  ``Scaling of cosmic string loops,''
  Phys.\ Rev.\  D {\bf 74}, 063527 (2006).

\bibitem{OV-06}
  K.~D.~Olum, V.~Vanchurin,
  ``Cosmic string loops in the expanding Universe,''
  Phys.\ Rev.\  {\bf D75}, 063521 (2007).


\bibitem{numerical-techniques-paper}
J.~J.~Blanco-Pillado, K.~D.~Olum and B.~Shlaer,
  ``A new parallel simulation technique,''
  arXiv:1011.4046 [physics.comp-ph].


\bibitem{Vitaly}
  V.~Vanchurin,
  ``Non-linear dynamics of cosmic strings with non-scaling loops,''
  Phys.\ Rev.\  {\bf D82}, 063503 (2010);
  V.~Vanchurin,
  ``Semi-scaling cosmic strings,''
  JCAP {\bf 1011}, 013 (2010).



\bibitem{VV}
  T.~Vachaspati and A.~Vilenkin,
  ``Formation And Evolution Of Cosmic Strings,''
  Phys.\ Rev.\  D {\bf 30}, 2036 (1984).

\bibitem{Talloires}
  ``Scaling distribution of large cosmic string loops,''
Talk presented by K.~D.~Olum at the conference, ``Challenges in
Theoretical Cosmology'', Talloires, France, (2009).


\bibitem{velocity-correction}
J.~J.~Blanco-Pillado, K.~D.~Olum and B.~Shlaer,
in preparation.

\bibitem{Bouchet-conference} 
F. Bouchet, ``High-resolution Simulations of Cosmic String Evolution:
Small Scale Structure and Loops'' in \emph{The formation and evolution
  of cosmic strings}, G. Gibbons, S. Hawking, and T. Vachaspati, eds.,
Cambridge University Press, Cambridge, England, (1990).

\end{thebibliography}
\end{document}